\documentclass[10pt, leqno]{article}
\usepackage[utf8]{inputenc}
\usepackage[T1]{fontenc}
\usepackage{amsmath}
\usepackage{amsfonts}
\usepackage{amssymb}
\usepackage[version=4]{mhchem}
\usepackage{stmaryrd}
\usepackage{graphicx}
\usepackage[export]{adjustbox}
\usepackage{csquotes}
\usepackage{hyperref}
\usepackage{authblk}
\usepackage{color}
\newcommand{\HRule}[1]{\rule{\linewidth}{#1}}

\title{ \normalsize \textsc{}
		\HRule{2pt} \\
		\LARGE \textbf{PT-Symmetry in One-Way Wormholes}
		\HRule{2.0pt}}

\author[1]{\small \textbf{Pascal Koiran}\thanks{\href{mailto:pascal.koiran@ens-lyon.fr}{pascal.koiran@ens-lyon.fr}}}
\author[2]{\small \textbf{Hicham Zejli}\thanks{\href{mailto:hicham.zejli@manaty.net}{hicham.zejli@manaty.net}}}
\author[2]{\small \textbf{J-P Levy}\thanks{\href{mailto:j.p.levy@manaty.net}{j.p.levy@manaty.net}}}
\author[3]{\small \textbf{Florent Margnat}\thanks{\href{mailto:florent.margnat@univ-poitiers.fr}{florent.margnat@univ-poitiers.fr}}}
\author[2]{\small \textbf{M-F Duval}\thanks{\href{mailto:marie-france.duval@manaty.net}{marie-france.duval@manaty.net}}}
\author[2]{\small \textbf{Hasnae Zejli}\thanks{\href{mailto:hasnae.zejli@manaty.net}{hasnae.zejli@manaty.net}}}

\affil[1]{École Normale Supérieure de Lyon}
\affil[2]{Manaty Research Group}
\affil[3]{Université de Poitiers}

\date{\today}

\begin{document}
\maketitle

\begin{abstract}
{In a recent paper, we studied a modified version of the Einstein-Rosen bridge.
This modified bridge is traversable and works as a one-way membrane: a particle on the first sheet falling toward the throat will reach it in finite time (in Eddington coordinates), and will continue its trajectory on the second sheet.
In this paper, we show that the particle undergoes a \textit{PT-symmetry} as it crosses the throat.
This could lead to observable effects thanks to an additional ingredient proposed by Einstein and Rosen: congruent points on the two sheets  are identified. We propose a bimetric model to realize this identification for our modified bridge.}


\end{abstract}

\newpage

\section{Solutions of Einstein’s equation reflecting different topologies} \label{sec:solutions}

{ We begin this paper by a review of some the work stemming from the discovery 
 by Schwarzschild of an exact solution to the Einstein field equations in vacuum. The work of Einstein and Rosen~\cite{[3]}  is of  particular importance for the present paper since we will be interested in the fate of a particle crossing an Einstein-Rosen bridge.  At first sight  this line of inquiry may look like  a dead end to some readers. 
  Indeed, the Einstein-Rosen bridge has often been
 presented as non-traversable in the literature. In Section 2 we point out that this conclusion is in fact based on an analysis 
 of the Kruskal-Szekeres extension, which as a geometric object is very different from an Einstein-Rosen bridge. 
 The main developments of the paper take place in Sections 3 and 4.  We show that a particle crossing the bridge undergoes a \textit{PT-symmetry}, and we discuss its physical significance. {We will in fact not work with the Einstein-Rosen bridge as
 defined in the seminal paper~\cite{[3]}, but with a modified version studied in~\cite{Koi}. One major reason for this modification
 is that, as explained below in Section~\ref{sec:solutions}, the bridge as defined in~\cite{[3]} is not properly glued at the throat
 in the following sense: as is well known,  infalling geodesics do not reach the wormhole throat for any finite value of the Scwharzschild time parameter $t$. 
 The construction in~\cite{Koi} is inspired by~\cite{Guend10,Guend17}
 and solves this problem.}\\

In 1916, Karl Schwarzschild successively published two papers (\cite{[9]},\cite{[10]}). The first one presented the construction of 
a solution to Einstein's equation in vacuum. 
{In its classical form under the signature \textit{(+ - - -)}, this is the well-known exterior Schwarzschild metric: 
\begin{equation}\label{eq:exterior}
\mathrm{d}s^2 = \left(1-\frac{\alpha}{r}\right) c^2 \mathrm{d}t^2 - \left(1-\frac{\alpha}{r}\right)^{-1} \mathrm{d}r^2 - r^2 (\mathrm{d}\theta^2 + \sin^2 \theta \mathrm{d}\varphi^2)
\end{equation}


As a complement to this exterior metric, he rapidly published an interior metric \cite{[10]}} describing the geometry inside a sphere filled with a fluid of constant density \(\rho_o\) and a solution to Einstein's equation with a second member. The conditions for connecting the two metrics (Continuity of geodesics) were ensured. The phenomena of the advance of Mercury's perihelion and gravitational lensing confirm this solution. K. Schwarzschild worked to ensure that the conditions governing these two metrics were in accordance with physical reality.\\

As an example, in the present day, neutron stars, owing to their staggering density and formidable mass, stand as natural cosmic laboratories, probing realms of density and gravity unreachable within terrestrial laboratories. Let us consider two distinct ways through which a neutron star might reach a state of physical criticality.\\

In a scenario where the star's density, $\rho_o$, remains constant, a characteristic radius $\hat{r}$ can be defined. Then, a physical criticality is reached when the star's radius is :
\begin{equation}\label{eq1}
R_{\text{cr}_{\phi}} = \sqrt{\frac{8}{9}} \hat{r} = \sqrt{\frac{c^2}{3\pi G \rho_o}}
\end{equation}
with 
\begin{equation}\label{eq2}
\hat{r} = \sqrt{\frac{3c^2}{8\pi G \rho_o}}
\end{equation}
Thus,
\begin{itemize}
    \item For the exterior metric, it was necessary that the radius of the star be less than \( \hat{r} \).
    \item As for the interior metric, the radius of the star had to be less than \( R_{\text{cr}_{\phi}} \) because a larger radius leads the pressure to rise to infinity at the center of the star.
\end{itemize}

Next, for massive stars, an imploding iron sphere can present a complex scenario. Assuming the sphere's mass $M$ is conserved during implosion, we must consider two important critical radius :\\

In the core part, the geometric criticality radius is given by the \textit{Schwarzschild Radius} which is :
\begin{equation}\label{eq3}
R_{\text{cr}_{\gamma}} = R_s = 2 \frac{GM}{c^2} 
\end{equation}

Outside of this mass, the physical critical radius is given by \ref{eq1}\\

With mass conservation expressed as \( M = \frac{4}{3}\pi R^3 \rho_o \), we can explore how the variable density $\rho_o$ during implosion impacts these critical radius.\\

Indeed, if physical criticality is reached during implosion, we have \( R = R_{\text{cr}_{\phi}} \).

Then, substituting the mass conservation equation into \ref{eq1}, we get :
\begin{equation}\label{eq4}
R = R_{\text{cr}_{\phi}} = 2.25 \frac{GM}{c^2} > R_{\text{cr}_{\gamma}}
\end{equation}

We can deduce that if the physical criticality is reached for a mass $M$, then it occurs before geometric criticality appears. \\

K. Schwarzschild also emphasized that the measurements pertained to conditions far exceeding what was understood within the framework of the astrophysical reality of his time.\\

It is also important to note that the topology of this geometric solution is built by connecting two bounded manifolds along their common boundary, a sphere \(S^2\) with an area of \(4 \pi R_o^2\).\footnote{\(R_o\) is the radius of the star.}\\

In 1916, Ludwig Flamm considered the external solution as potentially describing a geometric object. The concern was then an attempt to describe masses as a non-contractible region of space (\cite{[4]}).\\

In 1934, Richard Tolman was the first to consider a possible handling of the most general metric solution introducing a cross term \( \mathrm{d}r \, \mathrm{d}t \). However, for the sake of simplification, he immediately eliminated it using a simple change of variable (\cite{[14]}).\\

In 1935, Einstein and Rosen proposed, within the framework of a geometric modeling of particles, a non-contractible geometric structure, through the following coordinate change (\cite{[3]}):
\begin{equation}\label{eq5}
u^2 = r - 2m
\end{equation}
The metric solution then becomes:
\begin{equation}\label{eq6}
\mathrm{d}s^2 = \frac{u^2}{u^2 + 2m} \mathrm{d}t^2 - 4u^2(u^2 + 2m) \mathrm{d}u^2 - (u^2 + 2m)^2 (\mathrm{d}\theta^2 + \sin^2 \theta \mathrm{d}\phi^2)\\
\end{equation}

The authors thus obtain a non-contractible geometric structure, termed a \textit{"space bridge"}, where a closed surface of area \(4 \pi \alpha^2\), corresponding to the value \(u = 0\), connects two \textit{"sheets"}: one corresponding to the values of \(u\) from 0 to \(+\infty\) and the other from \(-\infty\) to 0. It is noteworthy that this metric is not Lorentzian at infinity\footnote{\label{chru} For this reason, the change of variables $r^2=\rho^2+4m^2$ was proposed  by Chruściel (\cite{[19]}, page 77) as an alternative to~(\ref{eq5}). See also the appendix of the present paper, where we propose an alternative to the change of variables from~\cite{[19]}.}. Although this metric, expressed in this new coordinate system, is regular, the authors point out that at the throat surface, its determinant becomes zero. 
In this geometric structure, two bounded semi-Riemannian sheets are distinguished, the first corresponding to \( u > 0 \) and the second to \( u < 0 \). It corresponds to their joining along their common boundary. The overall spacetime does not fit within the standard framework of semi-Riemannian geometry since it does not fulfill the requirement \( \det(g_{\mu\nu}) \neq 0 \) at the throat. As pointed out in \cite{[13]}, it does fit within the more general framework of singular semi-Riemannian geometry, which allows for degenerate metric tensors.\\


{The Einstein-Rosen bridge~(\ref{eq6}) 
satisfies the Einstein field equations in vacuum on both sheets $u>0$ and $u<0$. However, there is an issue with the field equations at the throat $u=0$ since  
\( \det(g_{\mu\nu})$ vanishes there, and this determinant appears in the denominator of the field equations. This issue was already recognized
by Einstein and Rosen, and their proposed solution was to work with a form of the field equations 
that is denominator-free (see equations (3a) in~\cite{[3]}, and the paragraph after (5a)). These modified field equations (called nowadays the "polynomial form of the field equations") are satisfied everywhere, including at the throat.\footnote{
We recall
in Appendix B how the modified field equations are obtained, and compare them to the standard form of the field equations in the context of the Einstein-Rosen bridge.}
{
Working with the standard form of the Einstein field equations, 
the authors of~\cite{Guend10,Guend17} discovered much later that a thin shell of "exotic matter" must be added at the throat for
the field equations to be satisfied. This feature is not apparent when one works with the polynomial form of the equations
and was thus overlooked by Einstein and Rosen. In particular,  the modified version of the bridge presented in Section~\ref{sec:3} has a metric (\ref{eq:bridge}) which is nondegenerate at the throat, and it becomes especially clear that 
the presence of a thin shell of exotic matter is necessary. For the details of this we refer again to~\cite{Guend10,Guend17}, 
where it is also explained how the presence of the thin shell can be seen on the original form of the bridge~(\ref{eq6}).}
}

As a spacetime, the Einstein-Rosen bridge~(\ref{eq6}) suffers from the problem that the time coordinate $t$ becomes infinite on the throat (since infall time to the throat is infinite in Schwarzschild coordinates). In~(\ref{eq6})  the 4-dimensional sheets $u>0$ and $u<0$ are therefore not properly glued at the throat.\footnote{Gluing the spatial (3-dimensional) parts of the two sheets does not raise any special difficulty, however.}
Namely, studying the passage of a particle from the sheet $u>0$ to $u<0$ would require 
to go through $t=\infty$, which is not a well defined part of the manifold.
We will see how to fix this problem in Section~\ref{sec:3}. }\\

In 1939, Oppenheimer and Snyder, capitalizing on the complete decoupling between proper time and the time experienced by a distant observer in {the exterior metric~(\ref{eq:exterior}) }
suggested using 
this solution  to describe the \textit{"freeze frame"} of the implosion of a massive star at the end of its life. By considering that the variable $t$ is identified with the proper time of a distant observer, it creates this \textit{"freeze frame"} pattern such as a collapse phenomenon whose duration, in proper time, measured in days, seems for a distant observer to unfold in infinite time (\cite{[8]}). This paper was considered as the foundation of the black hole model.\\

In 1960, Kruskal extended the geometric solution to encompass a contractible spacetime, organized around a central singularity corresponding to \( r = 0 \). The geodesics are extended for \( r < \alpha \). The black hole model (with spherical symmetry\footnote{In 1963, Roy Kerr constructed the stationary axisymmetric solution to Einstein's equation in vacuum. However, in this article, we limit ourselves to the interpretations of the stationary solution with spherical symmetry.}) then takes its definitive form as the implosion of a mass, in a brief moment, perceived as a \textit{"freeze-frame"} by a distant observer (\cite{[6]}). The Schwarzschild sphere is then termed the \textit{"event horizon"}.\\

In 1988, M. Morris and K. S. Thorne revisited this geometric interpretation by abandoning contractibility, not to attempt to obtain a geometric modeling of the solution, but to study the possibility of interstellar travel, through \textit{"wormholes"}, using the following metric (\cite{[7]}):
\begin{equation}\label{eq7}
\mathrm{d}s^2 = -c^2 \mathrm{d}t^2 + \mathrm{d}l^2 + (b_o^2 + l^2)(\mathrm{d}\theta^2 + \sin^2 \theta \mathrm{d}\phi^2)\\
\end{equation}

By focusing their study on the feasibility of interstellar travel, the authors highlight the enormous constraints associated with such geometry as well as its unstable and transient nature.

\begin{figure}[h]
\begin{center}
\includegraphics[max width=0.8\textwidth]{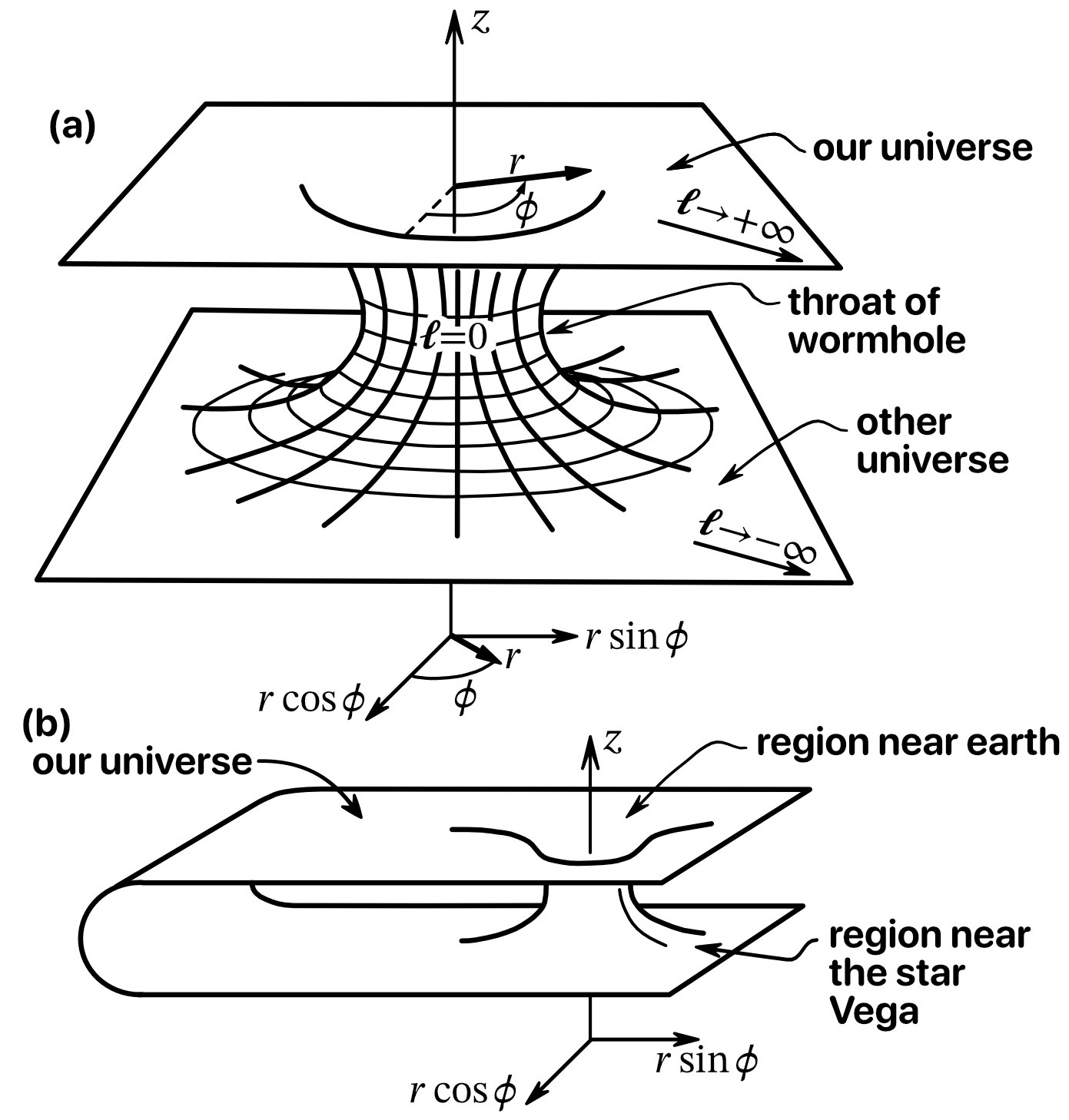}\\
\end{center}
\caption{Page 396 of the article by M. Morris and K.S. Thorne (1988)}
\label{fig:morris}
\end{figure}

\section{Distinction between the Kruskal-Szekeres Extension and the Einstein-Rosen Bridge}
\label{sec:distinct}

The Kruskal-Szekeres extension and the Einstein-Rosen bridge are two major constructions in the study of spacetime geometry around a wormhole. However, their geometric natures differ significantly.\\

The Kruskal-Szekeres spacetime is defined by a traditional \textit{semi-Riemannian manifold}, characterized by a non-degenerate metric at every point. This makes it consistent with the general framework of general relativity, where the metric's signature is homogeneous and does not vary {\cite{[MTW]}, \cite{[Wald]}}.\\

In contrast, the Einstein-Rosen spacetime has a degenerate metric at certain points, namely, at the bridge's throat. This characteristic places it in the class of \textit{singular semi-Riemannian manifolds} as defined by Ovidiu Stoica \cite{[13]}.\footnote{
It was recently proposed in~\cite{Guend24b} to regularize such spacetimes through a 
complexification based on a holomorphic extension of the theory of gravitation~\cite{Guend23}. We will not pursue this approach in the present paper since the main model that we study (namely, the "modified bridge" of Section~\ref{sec:3}) is nondegenerate at the throat.}
 This fundamental distinction shows that the Kruskal-Szekeres spacetime is not simply an extension of Einstein-Rosen but a fundamentally different construction.

{This geometric difference between the two spacetimes is also responsible for a physical difference. 
Indeed, as already mentioned in Section~\ref{sec:solutions}, for the field equations to be satisfied at the throat of the Einstein-Rosen bridge one needs to add a thin shell of exotic matter at the throat~\cite{Guend10,Guend17}.
  By contrast
the Kruskal-Szekeres  extension satisfies the Einstein field equations in vacuum, including at the event horizon. }\\


Thus, these two spacetimes cannot be considered versions of each other but rather two distinct interpretations of the geometry around a wormhole. This was already pointed out in several papers by Guendelman et al. Consider in particular \cite{Guend10}, where they write: 
\begin{quote}
[29] The nomenclature of ``Einstein-Rosen bridge'' in several standard textbooks (e.g. [15]) uses the Kruskal-Szekeres manifold. The latter notion of ``Einstein-Rosen bridge'' is not equivalent to the original construction in [14]. Namely, the two regions in Kruskal-Szekeres space-time corresponding to the outer Schwarzschild space-time region (r > 2m) and labeled (I) and (III) in [15] are generally \textit{disconnected} and share only a two-sphere (the angular part) as a common border (U = 0, V = 0 in Kruskal-Szekeres coordinates), whereas in the original Einstein-Rosen ``bridge'' construction the boundary between the two identical copies of the outer Schwarzschild space-time region (r > 2m) is a three-dimensional hypersurface (r = 2m).
\end{quote}


We can also cite two other papers whose authors make the same observation regarding the Kruskal-Szekeres extension's inadequacy in properly analyzing the Einstein-Rosen bridges: that of Guendelman et al. \cite{Guend17} and that of Poplawski \cite{[18]}. Indeed, to distinguish these spacetimes, Poplawski uses the terms \textit{"Schwarzschild bridge"} and \textit{"Einstein-Rosen bridge"}.\\

For all these reasons, we will {\em not} work with the Kruskal-Szekeres extension in this paper. We note in particular that the
common claim~\cite{[FW],[MTW]} that the Einstein-Rosen bridge is not traversable is actually based on an analysis of the Kruskal-Szekeres extension; but, as pointed out in~\cite{Guend17,Koi}, the original Einstein-Rosen bridge~\cite{[3]} is in fact traversable.

{\section{The modified bridge and its symmetries}
\label{sec:3}
}
{In this section we study the symmetries of a modified version of the original Einstein-Rosen bridge~\cite{[3]}}.
{As recalled in Section~\ref{sec:solutions}, Einstein and Rosen defined their "bridge" from the change of variables $r=\alpha+u^2$ in~(\ref{eq:exterior}). 
The definition of the modified bridge is based on the idea from~\cite{Guend10,Guend17} to work instead with the change of variables $r=\alpha+|\eta|$ where $\eta \in \mathbb{R}$ is a new radial parameter. As shown in~\cite{Guend10,Guend17}, the resulting spacetime satisfies the 
Einstein field equations, 
including at the throat $\eta=0$, if some "exotic matter" (a lightlike membrane) is added at the 
throat. 

As we now explain, the modified bridge studied in this section is obtained by combining the change of variables 
$r=\alpha+|\eta|$ with Eddington's change of variables for the time parameter.}\\

{\subsection{\textit{PT-symmetry}}
\label{sec:PT}


  Eddington~\cite{[2]}  introduced his change of variables  
\begin{equation}\label{eq:tE+}
{t}^{+}_E={t}+\frac{\alpha}{{c}} \ln \left|\frac{ r }{\alpha}-1\right|
\end{equation}
with the aim of eliminating the coordinate singularity at the Schwarzschild surface in \( r = \alpha \). The metric becomes:
\begin{equation}\label{eq:E+metric}
\mathrm{d}s^{2}=\left(1-\frac{\alpha}{ r }\right) {c}^{2} \mathrm{d}{t^{+}_E}^{2}-\left(1+\frac{\alpha}{ r }\right) \mathrm{d}r^{2}-\frac{2 \alpha {c}}{ r } \mathrm{d}r \mathrm{d}t^{+}_E- r ^{2}\left(\mathrm{d} \theta^{2}+\sin ^{2} \theta \mathrm{d} \varphi^{2}\right)
\end{equation}

We know that under these conditions 
 free fall time is finite in Eddington coordinates, i.e., a massive infalling particle will reach the surface $r=\alpha$ for a finite value of $t_E^+$~\cite{Koi}. It is  however  well-known that the surface $r=\alpha$ is not reached for any finite value of the Schwarzschild 
 time parameter $t$.\\

By contrast,  escape time {for $t_E^+$} remains infinite. The metric for which the escape time is finite will be obtained by performing this change of variable:
\begin{equation}\label{eq11}
{t}^{-}_E={t}-\frac{\alpha}{{c}} \ln \left|\frac{ r }{\alpha}-1\right|
\end{equation}
In this case, the metric becomes:
\begin{equation} \label{eq:E-metric}
\mathrm{d}s^{2}=\left(1-\frac{\alpha}{ r }\right) {c}^{2} \mathrm{d}{t^{-}_E}^{2}-\left(1+\frac{\alpha}{ r }\right) \mathrm{d}r^{2}+\frac{2 \alpha {c}}{ r } \mathrm{d}r \mathrm{d}t^{-}_E- r ^{2}\left(\mathrm{d} \theta^{2}+\sin ^{2} \theta \mathrm{d} \varphi^{2}\right)
\end{equation}
The modified bridge studied in~\cite{Koi} combines the change of variables $r=\alpha+|\eta|$ with~(\ref{eq:tE+}), i.e.,
we work with the new time parameter $t'=t+\frac{\alpha}{{c}} \ln \left|\frac{ \eta }{\alpha}\right|.$\\

Thus, the metric becomes:
\begin{equation}\label{eq:bridge}
\mathrm{d}s^2=\frac{|\eta|}{\alpha+|\eta|}c^2\mathrm{d}t'^2-\frac{2\alpha+|\eta|}{\alpha+|\eta|}{\mathrm{d}\eta^2}-\frac{2\alpha c}{\alpha+|\eta|}\mathrm{d}\eta\ \mathrm{d}t'
 -(\alpha+|\eta|)^{2}\left(\mathrm{d} \theta^{2}+\sin ^{2} \theta \mathrm{d} \varphi^{2}\right).
\end{equation}
This line element already appears in the appendix of~\cite{Guend10} in slightly different notation. It describes a spacetime made of two sheets connected at the throat $\eta=0$. The sheet $\eta>0$ is equipped with the ingoing Eddington metric~(\ref{eq:E+metric}) and the sheet $\eta<0$ is equipped with the outgoing metric~(\ref{eq:E-metric}).  
As pointed out in~\cite{Koi}, an infalling particle beginning its trajectory in the region $\eta>0$ will reach the throat $\eta=0$ for a finite value {$t'_1$} of the Eddington time parameter $t'$, and will then continue in the region $\eta<0$ {for $t'>t'_1$. }This resolves the gluing problem 
that was mentioned in Section 1 for the original version of the Einstein-Rosen bridge (recall that the throat is reached for 
$t=\infty$ with the bridge as defined in~\cite{[3]}). {Note also that unlike the Einstein-Rosen metric~(\ref{eq6}), metric~(\ref{eq:bridge}) is nondegenerate at the throat.}\\

The line element~(\ref{eq:bridge}) is invariant under the joint transformations $\eta \mapsto - \eta, t' \mapsto -t'$. 
The physical significance of this symmetry will be discussed in Section~\ref{sec:3.3}.
Note that the line element~(\ref{eq6}) has a similar symmetry, and in fact it is {\em more} symmetric since it is invariant
under each of the two transformations $u \mapsto -u$ and $t \mapsto -t$. This extra symmetry is due to the absence in~(\ref{eq6})  of 
a cross term such as the term $\mathrm{d}\eta \mathrm{d}t'$ in~(\ref{eq:bridge}).

}

\begin{figure}[h]
\begin{center}
\includegraphics[max width=\textwidth]{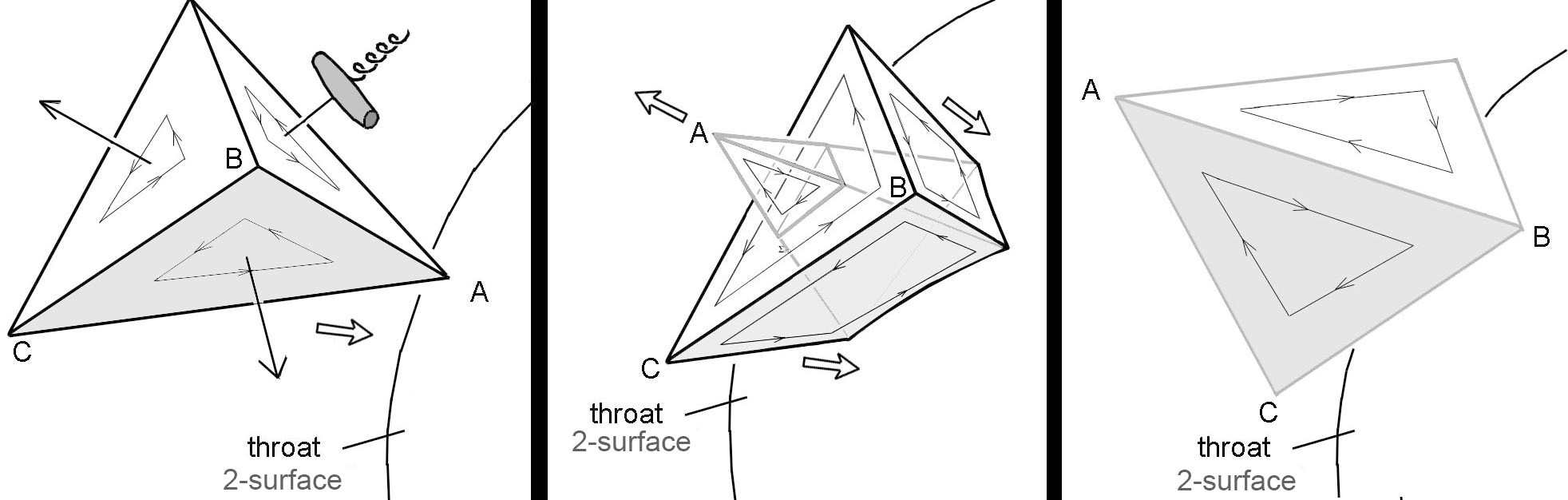}\\
\end{center}
\caption{Inversion of space when crossing the \textit{"space bridge"}}
\label{fig:inversion}
\end{figure}

{
\subsection{Change of orientation}
\label{sec:3.2}

In general, we expect a \textit{P-symmetry} or \textit{PT-symmetry} to be associated to a change of orientation. In this section we confirm that this is indeed the case by taking a closer look at the geometry of the modified bridge~(\ref{eq:bridge}) in the vicinity of the throat $\eta=0$.}
In this representation, the radial geodesics of the first sheet are orthogonal to the tangent plane at the \textit{"space bridge"} when they reach it. These same geodesics, emerging in the second sheet, are also orthogonal to this same tangent plane. Let's now consider four points forming a tetrahedron, which converge towards the \textit{"space bridge"} along radial trajectories. We can set a 3D orientation by defining a direction of traversal of the points on each of the equilateral triangles forming the tetrahedron. With respect to the coordinate {\( r=\alpha+|\eta| \)}, it seems as if these points bounce off a rigid surface, leading to an inversion of the orientation of the tetrahedron. The upstream and downstream tetrahedra then become \textit{enantiomorphic} (Figure~\ref{fig:inversion}).

The change of orientation is already visible in the simplified 2-dimensional representation of a wormhole in Figure~\ref{fig:morris}. Let us look at this figure from above, and imagine a triangle gliding on the surface of the top sheet toward the throat. After crossing the throat, the triangle starts gliding on the bottom sheet and we now see it upside now from our position above the top sheet. From our point of view, its orientation has therefore changed. The physical meaning of this change of orientation will be discussed in Section \ref{sec:3.3}.\\

As a geometric structure, metric~(\ref{eq:bridge}) 
represents a \emph{"bridge"} connecting two \emph{PT-symmetric} semi-Riemannian spaces.\\

The element of this 2D-surface is 
given by:
\begin{equation}\label{eq16}
\sqrt{|\det(g_{\mu \nu})|} = \sqrt{|g_{\theta \theta} g_{\phi \phi}|} = \alpha^2 \sin(\theta)
\end{equation}

As this metric describes a 2D-surface sphere (like a sphere of constant radius in a 4D spacetime), then the differential area element is given by :
\begin{equation}\label{eq17}
\mathrm{d}A = \sqrt{|\det(g_{\mu \nu})|} \mathrm{d}\theta \mathrm{d}\phi = \alpha^2 \sin(\theta) \mathrm{d}\theta \mathrm{d}\phi
\end{equation}

To find the minimal area of this \textit{"space bridge"}, we must integrate this area element over all possible angles :
\begin{equation}\label{eq18}
A = \int_0^{2\pi} \int_0^{\pi} \alpha^2 \sin(\theta) \mathrm{d}\theta \mathrm{d}\phi = 4\pi \alpha^2
\end{equation}

It is therefore non-contractible with a minimal area of \(4\pi \alpha^2\).

\subsection{Identification of the two sheets}
\label{sec:3.3}

In Section \ref{sec:3.2} we have described the change of orientation of a tetrahedron crossing the wormhole throat in Figure~\ref{fig:inversion}, and of a triangle crossing the throat { in Figure~\ref{fig:morris}}. The change of orientation of the triangle is only visible for a person looking at Figure~\ref{fig:morris} in its entirety. Therefore, it does not correspond to any physically observable phenomenon since any physical observer must be located on one of the two sheets and cannot see directly the other sheet. The situation is the same in Figure~\ref{fig:inversion} : The middle picture represents the situation from a point of view where we could look simultaneously at the two sides of the wormhole (B and C have not reached the throat yet, while A has already crossed it and emerges on the other side). This is again impossible for a physical observer: it seems that the 
{\textit{PT-symmetry}} as described so far does not correspond to any physically observable phenomenon. We can however give it a real physical meaning with an additional ingredient due to Einstein and Rosen \cite{[3]}. \\

Recall that their motivation was not to investigate interstellar travel as in Figure~\ref{fig:morris}, but to describe elementary particles by solutions to the equations of general relativity. Quoting from the abstract of their paper: \textit{"These solutions involve the mathematical representation of physical space by a space of two identical sheets, a particle being represented by a "bridge" connecting these sheets."} Einstein and Rosen also suggest that the multi-particle problem might be studied by similar methods, but this work is not carried out in their paper.\\

Quoting again from \cite{[3]} : \textit{"If several particles are present, this case corresponds to finding a solution without singularities of the modified Eqs. (3a), the solution representing a space with two congruent sheets connected by several discrete "bridges.""} 
From their point of view, two points in the mathematical representation (\ref{eq6}) with identical values of \(\theta, \phi\) but opposite values of \(u\) therefore correspond to two points in physical space with the same value of \(r\) (\(r = u^2 +2m\)). If we make the same identification of points with opposite values of \(u\), the situation represented in the middle picture of Figure~\ref{fig:inversion} can be seen by a physical observer. 
The 
{change of orientation} described in Section~\ref{sec:3.2} now has a real physical meaning. We will elaborate on the interpretation of the combined \emph{PT-symmetry} in 
Section~\ref{sec:inter}. In Section~\ref{sec:bimetric} we present an accurate mathematical model of the identification of the two sheets for the modified bridge described in Section~\ref{sec:PT}.
It turns out to be a bimetric model.

\subsection{Interpretation of the \textit{PT-symmetry}}
\label{sec:inter} 

{\textit{PT-symmetry} can be viewed as a \textit{P-symmetry} followed by a \textit{T-symmetry}.}\\

In the literature, the inversion of the time coordinate has been analyzed in various ways. In particular:\\

\begin{itemize}
\item It was analyzed through the dynamic group theory of J-M Souriau (\cite{[11]},\cite{[12]}), and was shown to result in an inversion of energy.  Consequently, time reversal transforms every motion of a particle of mass $m$ into a motion of a particle of mass $-m$ (\cite{[12]}, page 191). On page 192 of the same book, the author offers an alternative analysis which avoids negative masses. Souriau points out that these alternatives should  be judged according to their ability to explain experiments.

\item Feynman has offered an interpretation of antimatter as ordinary matter traveling backward in time. 

{
\item  In the context of string theory,   a  proposal similar to Feynman's can  be found in~\cite{Guend24}. 
Indeed, quoting from~\cite{Guend24}:  \textit{"The antistrings are realized when a scalar time that defines the modified measure runs in the opposite direction to the world sheet time. For strings
with positive tension, both times run in the same direction. The situation resembles the situation in Relativistic Quantum Mechanics with positive and negative
energies, proper time of particles running forward with respect of coordinate
time, while for antiparticles proper time runs opposite of coordinate time."}

}

\item It is known from theoretical analysis (the \textit{CPT} theorem) and from experiments that elementary particles obey physical 
laws that are invariant under \textit{CPT-symmetry}.
\end{itemize}

{The \textit{PT-symmetry} uncovered in Section \ref{sec:3} can be viewed as a \textit{CPT-symmetry} followed by a \textit{C-symmetry} (inversion of electric charge). This suggests that the \textit{PT-symmetry} might lead to observable effects where particles and antiparticles are connected through the bridge. If the second sheet already contains ordinary matter, it could interact with the antimatter coming from the first sheet, constituting a potential source of energy. This provides a profound physical interpretation of the \textit{PT-symmetry} in our geometric construction beyond the simple choice of coordinates.}

{\section{The Bimetric Bridge}
\label{sec:bimetric}


In  Section~\ref{sec:3.3} we explained that according to~\cite{[3]},  two points in the mathematical representation (\ref{eq6}) with identical values of \(\theta, \phi\) but opposite values of \(u\)  correspond to two points in physical space with the same value of \(r\) (\(r = u^2 + 2m\)). If we identify in~(\ref{eq6}) two points with opposite values of $u$, it seems that 
we are just left with a single sheet carrying up to a change of variables the Schwarzschild solution in} 
{the region $r>\alpha$. The "throat" $u=0$ (or $r=\alpha$, in Schwarzschild coordinates) therefore appears as a limit of space rather than a gateway to a second sheet.\\

For the modified bridge studied in Section~\ref{sec:PT}, the situation is more interesting because the two sheets
carry different metrics (the ingoing and outgoing Eddington metrics).
After identification of two points with opposite values of~$\eta$ in~(\ref{eq:bridge}), we are again left with a single sheet
but it is equipped with the two metrics:
\begin{equation}\label{eq:ds+}
\mathrm{d}s_+^{2}=\left(1-\frac{\alpha}{ r }\right) {c}^{2} \mathrm{d}{t}^{2}-\left(1+\frac{\alpha}{ r }\right) \mathrm{d}r^{2}-\frac{2 \alpha {c}}{ r } \mathrm{d}r \mathrm{d}t- r ^{2}\left(\mathrm{d} \theta^{2}+\sin ^{2} \theta \mathrm{d} \varphi^{2}\right)
\end{equation}

\begin{equation} \label{eq:ds-}
\mathrm{d}s_-^{2}=\left(1-\frac{\alpha}{ r }\right) {c}^{2} \mathrm{d}{t}^{2}-\left(1+\frac{\alpha}{ r }\right) \mathrm{d}r^{2}+\frac{2 \alpha {c}}{ r } \mathrm{d}r \mathrm{d}t- r ^{2}\left(\mathrm{d} \theta^{2}+\sin ^{2} \theta \mathrm{d} \varphi^{2}\right)
\end{equation}

 We therefore obtain a bimetric model. A particle that is infalling according to the first metric will reach the throat $r=\alpha$
 for a finite value of $t$ (say, $t=t_0$). For $t>t_0$ the second (outgoing) metric takes over. The particle will effectively 
 {\em seem} to go back in time (it retraces its steps) since~(\ref{eq:ds-}) is obtained from~(\ref{eq:ds+}) by the transformation
 $t \mapsto -t$. 
 Namely, for $\tau >0$ the particle's position at time $t_0+\tau$  will be the same as at time $t_0-\tau$ (when it was governed by the first metric).
 This is consistent with the \textit{PT-symmetry} uncovered in Section~\ref{sec:PT}.
 Additionally, the change of orientation highlighted in Section~\ref{sec:3.2} remains observable in this bimetric model. Indeed, as it rebounds from the throat, the tetrahedron depicted in Figure \ref{fig:inversion} undergoes a reflection relative to the tangent plane at the throat, leading to a change in orientation.
 {
 
\subsubsection*{Comparison with Hossenfelder's bimetric theory} 
 
 A \textit{"bimetric theory with exchange symmetry"} was proposed by Hossenfelder in~\cite{Ho}. There are two types of matter in this theory, which can be viewed as matter of \textit{"positive mass"} and 
 \textit{"negative mass"}. Each type of matter follows  the geodesics of  its own metric. In  Hossenfelder's theory, metric~(\ref{eq:ds+}) describes the movement of positive masses in a field created by a point of positive mass since her theory reduces to General Relativity in this case. We therefore obtain the ordinary Schwarszchild metric (see equation~(36) in~\cite{Ho}), or~({\ref{eq:ds+}) 
 in the ingoing Eddington coordinates. It is therefore natural to ask whether~(\ref{eq:ds-}) might describe the movement of a particle of {\textit{negative mass}} in the field created by the same \textit{positive mass} as in~(\ref{eq:ds+}). 
 The answer to this question is negative because in Hossenfelder's theory, the corresponding metric is obtained from the ordinary  Schwarszchild metric~(\ref{eq:exterior}) by the transformation $\alpha \mapsto -\alpha$ (equation~(37) in~\cite{Ho}). If we apply Eddington's change of variables to the resulting metric, we do not obtain anything like~(\ref{eq:ds-}).\footnote{As pointed out in Section~V of~\cite{Ho}, "changing to one of the more well-behaved systems with e.g. in-/outgoing Eddington-Finkelstein coordinates will be a nice transformation for the usual metric {\bf g}, but completely mess up the other metric $\underline{\bf h}$."} }

}
}

\section{Conclusion}

{In this paper, we have continued the study of the modified Einstein-Rosen bridge from~\cite{Koi}. We have recalled that the modified bridge is traversable, behaves as a one-way membrane and solves a \textit{"gluing problem"} 
from which the original version of  the bridge~\cite{[3]} suffered. {Moreover, the metric of this modified bridge is nondegenerate at the throat.}
The main contribution of the present paper lies in the study of the symmetries of the modified bridge.
We have pointed out that this structure is made of two  
\emph{PT-symmetric} enantiomorphic semi-Riemannian spaces (the two "sheets") with a Lorentzian metric at infinity, connected through the wormhole throat. The \emph{PT-symmetry} may seem like an artifact of the choice of coordinates, devoid of any physical meaning. However,   we have shown that this \emph{PT-symmetry} leads to observable effects thanks to an additional ingredient due to Einstein and Rosen. In~\cite{[3]} they suggested to represent a point in physical space by a pair of congruent points, one on each of the two sheets. For the original version of the bridge, this identification only seems to leave us with one of the two sheets (i.e., with the Schwarzschild solution for $r > \alpha$). The situation is different for the modified bridge since each sheet is 
equipped with a different metric. As a result, we obtain after identification a bimetric model. 
Finally, we have shown that this bimetric model is {\em not} consistent with Hossenfelder's bimetric theory~\cite{Ho}.}

\appendix

\section*{Appendix A: Another Representation of this Geometry}

{Instead of the change of variable $r=\alpha+|\eta|$, which yields~(\ref{eq:bridge}), one can consider the
smooth change of variable:
\begin{equation}\label{eq19}
r = \alpha \left(1 + \text{log ch}(\rho) \right).
\end{equation} 

By performing this  change of variable in the ingoing Eddington metric~(\ref{eq:E+metric}) and
the outgoing metric~(\ref{eq:E-metric}) we obtain:
\begin{equation}\label{eq20}
\begin{split}
\mathrm{d}s^{2} = &\left(\frac{\text{log cosh}(\rho)}{1 + \text{log cosh}(\rho)}\right) c^{2} \mathrm{d}{t}^{2} - \left(\frac{2+\text{log cosh}(\rho)}{1 + \text{log cosh}(\rho)}\right) \alpha^{2} \tanh^{2}(\rho) \mathrm{d}\rho^{2}\\
&- 2c \alpha \left(\frac{\tanh(\rho)}{1 + \text{log cosh}(\rho)}\right) \mathrm{d}\rho \mathrm{d}t -\alpha^{2}(1 + \text{log cosh}(\rho))^{2} (\mathrm{d}\theta^{2} + \sin^{2}\theta \mathrm{d}\varphi^{2})
\end{split}
\end{equation}
\begin{equation}\label{eq21}
\begin{split}
\mathrm{d}s^{2} = &\left(\frac{\text{log cosh}(\rho)}{1 + \text{log cosh}(\rho)}\right) c^{2} \mathrm{d}t^{2} - \left(\frac{2+\text{log cosh}(\rho)}{1 + \text{log cosh}(\rho)}\right) \alpha^{2} \tanh^{2}(\rho) \mathrm{d}\rho^{2}\\
&+ 2c \alpha \left(\frac{\tanh(\rho)}{1 + \text{log cosh}(\rho)}\right) \mathrm{d}\rho \mathrm{d}t- \alpha^{2}(1 + \text{log cosh}(\rho))^{2} (\mathrm{d}\theta^{2} + \sin^{2}\theta \mathrm{d}\varphi^{2})
\end{split}
\end{equation}}

Those metrics 
structure two sheets corresponding to $\rho$ varying respectively from 0 to \(+\infty\) and \(-\infty\) to 0. {We note that that these two metrics are properly glued at the throat $\rho=0$ since they both reduce to the same metric
$$ds^2= -\alpha^{2} (\mathrm{d}\theta^{2} + \sin^{2}\theta \mathrm{d}\varphi^{2})$$
} 
{on the throat. We can observe the same PT-symmetry as in Section~\ref{sec:PT}, i.e., from~(\ref{eq20})
 the joint transformations $\rho \mapsto -\rho$, $t \mapsto -t$ yield~(\ref{eq21}).}\\
 
 On the \emph{"space bridge"} for \(\rho=0\), the components \( g_{tt} \), {$g_{t\rho}$} and  \( g_{\rho\rho} \) of the metric tensor disappear, leaving only the last two spatial components \( g_{\theta\theta} \) and \( g_{\phi\phi} \), which are:
\begin{equation}\label{eq22}
g_{\mu\nu} = \begin{pmatrix}
0 & 0 & 0 & 0 \\
0 & 0 & 0 & 0 \\
0 & 0 & -\alpha^2 & 0 \\
0 & 0 & 0 & -\alpha^2 \sin^2 \theta \\
\end{pmatrix}
\end{equation}

On this particular coordinate system, we can infer that its determinant is zero. {Recall that the metric was also degenerate at the throat in the original version of the Einstein-Rosen bridge~(\ref{eq6}), but nondegenerate for the "modified bridge"~(\ref{eq:bridge}). }\\

One nice property of the change of variables~(\ref{eq19}) is that the two resulting metrics are explicitly Lorentzian as $|\rho| \rightarrow +\infty$.  This was also the motivation for the change of variables $r^2=\rho^2+4m^2$ proposed by Chruściel~\cite{[19]} and already  mentioned in footnote~\ref{chru}. {As we have seen, the nonsmooth change of variable $r=\alpha+|\eta|$ also leads to an explicitly Lorentzian metric.}

{
\section*{Appendix B: Polynomial form of the Einstein field equations}

The standard form of the field equations is
\begin{equation} \label{eq:EFE}
G_{\mu \nu} = - T_{\mu \nu}
\end{equation}
where $G_{\mu \nu}$ is the Einstein tensor and $T_{\mu \nu}$ is the stress-energy tensor. We are using here and throughout the paper the same sign convention as in~\cite{[3]}, and we have set the Einstein gravitational constant to 1 also as in~\cite{[3]}.
The determinant  of the metric tensor $g_{\mu \nu}$ appears in the denominator of the left-hand side.
As pointed out in~\cite{[3]} one can get rid of the denominators by multiplying the field equations 
by a suitable power of $\det(g)$. The new equations
\begin{equation} \label{eq:poly}
G^*_{\mu \nu} = - T^*_{\mu \nu}
\end{equation}
are polynomial equations in the metric tensor and its first and second order derivatives.
The polynomial form~(\ref{eq:poly}) is of course equivalent to~(\ref{eq:EFE}) except possibly when 
$\det(g)$ vanishes and the issue of a division by 0 in~(\ref{eq:EFE}) arises. This is the issue
that Einstein and Rosen had to face in the study of their "bridge", and they chose to 
work with the polynomial form of the equations to avoid a division by 0 at the bridge.
More precisely, since they were looking only for vacuum solution they 
worked with   the equations
\begin{equation} \label{eq:vacuum}
G^*_{\mu \nu} = 0.
\end{equation}
Then they argued that these equations are satisfied by their metric solution~(\ref{eq6}) everywhere including at the throat.
This indeed follows from the fact that~(\ref{eq:vacuum}) is satisfied outside of the throat (where the standard and polynomial
form of the field equations are equivalent), and from a continuity argument: the metric tensor~(\ref{eq6}) is infinitely differentiable (for this argument to go through we only need it to be twice differentiable, with continuous derivatives).

Although the above argument from~\cite{[3]} is mathematically correct, one can argue on physical grounds that it is preferable
to work with the standard form of the field equations as in~\cite{Guend10,Guend17}. One is then naturally led to the concept
of a lightlike membrane sitting at the throat. This possibility is not at all apparent with the polynomial form of the equations.
Indeed, the right-hand side of~(\ref{eq:poly}) vanishes where $\det(g)=0$ since it is obtained from the  $T_{\mu \nu}$ 
by multiplication by a power of $\det(g)$. This multiplication effectively "hides" the presence of the lightlike membrane.
Finally, we note that (as already pointed out in  Section~\ref{sec:solutions}), the presence of a lightlike membrane
is especially apparent in the modified bridge of Section 3, which is nondegenerate at the throat (and has a  metric tensor 
which is not differentiable at the throat). Since the lightlike membrane is present in the modified bridge~(\ref{eq:bridge}), it should also be present in the original form~(\ref{eq6})  of the bridge (one version of the bridge  is indeed obtained from the other by a change of variable).
}

\end{document}